\documentclass[floats,floatfix,showpacs,amssymb,prd,twocolumn,superscriptaddress,nofootinbib]{revtex4-2}

\usepackage{graphicx}
\usepackage{dcolumn}
\usepackage{bm}
\usepackage{amsmath}
\usepackage[colorlinks=true,linkcolor=blue,citecolor=blue,urlcolor=blue]{hyperref}
\usepackage{dsfont}

\usepackage{color}
\usepackage{ulem}

\newcommand{\bfx}{{\bf x}}
\newcommand{\bfu}{{\bf u}}

\newcommand{\bfP}{{\bf P}}
\newcommand{\bfv}{{\bf v}}

\newcommand{\bfnabla}{\bm{\nabla}}
\newcommand\comment[1]{}

\makeatletter
\def\doauthor#1#2#3{%
  \ignorespaces#1\unskip
  \begingroup
   #3%
  \@if@empty{#2}{\@listcomma\endgroup{}{}}{\endgroup{\comma@space}{}\frontmatter@footnote{#2}}%
  \space \@listand
}%
\makeatother

\makeatletter
\def\@ssect@ltx#1#2#3#4#5#6[#7]#8{%
  \def\H@svsec{\phantomsection}%
  \@tempskipa #5\relax
  \@ifdim{\@tempskipa>\z@}{%
    \begingroup
      \interlinepenalty \@M
      #6{%
       \@ifundefined{@hangfroms@#1}{\@hang@froms}{\csname @hangfroms@#1\endcsname}%
       {\hskip#3\relax\H@svsec}{#8}%
      }%
      \@@par
    \endgroup
    \@ifundefined{#1smark}{\@gobble}{\csname #1smark\endcsname}{#7}%
  }{%
    \def\@svsechd{%
      #6{%
       \@ifundefined{@runin@tos@#1}{\@runin@tos}{\csname @runin@tos@#1\endcsname}%
       {\hskip#3\relax\H@svsec}{#8}%
      }%
      \@ifundefined{#1smark}{\@gobble}{\csname #1smark\endcsname}{#7}%
      \addcontentsline{toc}{#1}{\protect\numberline{}#8}%
    }%
  }%
  \@xsect{#5}%
}%
\makeatother

\begin{document}


\title{Rapid Methods for Modeling Overdensities of \\ Massive Neutrinos and Other Non-Cold Relics}

\author{Keduse Worku$^{\mathds{W},}$}
\affiliation{William H.\ Miller III Department of Physics \& Astronomy, Johns Hopkins University, 3400 N.\ Charles St., Baltimore, MD 21218, USA}

\author{Nashwan Sabti$^{\mathds{S},}$}
\affiliation{William H.\ Miller III Department of Physics \& Astronomy, Johns Hopkins University, 3400 N.\ Charles St., Baltimore, MD 21218, USA}

\author{Marc Kamionkowski$^{\mathds{K},}$}
\affiliation{William H.\ Miller III Department of Physics \& Astronomy, Johns Hopkins University, 3400 N.\ Charles St., Baltimore, MD 21218, USA}


\def\thefootnote{$\mathds{W}$\hspace{-1.9pt}}\footnotetext{\href{mailto:kworku2@jhu.edu}{kworku2@jhu.edu}}
\def\thefootnote{$\mathds{S}$\hspace{0pt}}\footnotetext{\href{mailto:nash.sabti@gmail.com}{nash.sabti@gmail.com}}
\def\thefootnote{$\mathds{K}$\hspace{-0.9pt}}\footnotetext{\href{mailto:kamion@jhu.edu}{kamion@jhu.edu}}

\setcounter{footnote}{0}
\def\thefootnote{\arabic{footnote}}

\begin{abstract}
\noindent Recent work has highlighted the potentially detectable gravitational-lensing effect of neutrino halos on cosmic-microwave-background (CMB) fluctuations with upcoming instruments like SO, CMB-S4, and CMB-HD. Accurate modeling of neutrino-halo density profiles are essential for making theory predictions of their cosmological effects. Yet, they are computationally intensive, particularly in the nonlinear regime. In this work, we present an efficient numerical framework for computing neutrino profiles based on $N$-1--body simulations within a flat FRW Universe. Our approach enables highly parallelized, rapid calculations of neutrino trajectories near spherically symmetric dark-matter halos, delivering results within seconds. In addition to neutrinos, we demonstrate an application to model the clustering of other non-cold relics, such as a freeze-in dark-matter component. The framework is flexible in its definitions of cosmological and dark-matter-halo parameters, which can be particularly valuable for rapid-scanning tasks. It can also seamlessly incorporate new physics, as we demonstrate with examples of a time-varying gravitational constant and nonstandard neutrino phase-space distributions. 
    \\\\
    {\centering\noindent \href{https://github.com/NNSSA/Cheetah}{\raisebox{-1pt}{\includegraphics[width=9pt]{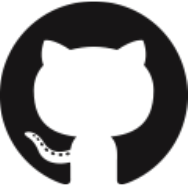}}}\hspace{2pt} Code and plotting scripts can be found \href{https://github.com/NNSSA/Cheetah}{here}.}
\end{abstract}

\maketitle

\section{Introduction}
The $\Lambda$CDM standard model of cosmology predicts the existence of a cosmic neutrino background (C$\nu$B) that decoupled within the first few seconds after the Big Bang, subsequently evolving as a free-streaming species~\cite{Gershtein:1966gg}. Detecting these relic neutrinos remains one of the foremost challenges in modern cosmology and particle physics. Due to their extremely weakly-interacting nature, they have thus far evaded experimental detection~\cite{Cocco:2008nka}. Nevertheless, significant efforts are being made to detect them both directly and indirectly~\cite{Alvey:2021xmq, PTOLEMY:2022ldz, PTOLEMY:2019hkd, Langacker:1982ih, Duda:2001hd, Eberle:2004ua, Yoshimura:2014hfa, Domcke:2017aqj, Alonso:2018dxy, Shergold:2021evs, Bauer:2021uyj, Bauer:2022lri}.

Despite limited direct evidence, a substantial amount is already known about relic neutrinos. At the background level, this includes constraints on their relativistic energy density from big-bang-nucleosynthesis (BBN) and cosmic-microwave-background (CMB) observations through the effective number of relativistic degrees of freedom $N_\mathrm{eff}$~\cite{Fields:2019pfx,Planck:2018vyg}. At the linear level, relic neutrinos suppress small-scale structure formation due to their non-zero masses, as has been observed in both CMB and galaxy surveys~\cite{Lesgourgues:2006nd, eBOSS:2015jyv, eBOSS:2020yzd, DES:2021wwk}, allowing for constraints on the sum of neutrino masses~\cite{Planck:2018vyg, Shao:2024mag}. Within the $\Lambda$CDM framework, we also expect the relic neutrino background to follow a Fermi-Dirac distribution with a temperature of $1.95(1+z)\,\mathrm{K}$, which can be used to set an upper bound on the sum of neutrino masses of $\sum m_\nu < 0.12\,\mathrm{eV}$~\cite{Planck:2018vyg}. However, the CMB remains largely insensitive to the exact shape of the neutrino phase-space distribution~\cite{Cuoco:2005qr, Fields:2019pfx, Oldengott:2019lke, Alvey:2021xmq}, making this upper limit model-dependent and opening the possibility for it to be relaxed to as high as $3\,\mathrm{eV}$ in scenarios deviating from $\Lambda$CDM~\cite{Alvey:2021xmq}. 

The existence of non-zero neutrino masses has spurred a fairly significant body of work on the clustering of relic neutrinos around dark-matter halos~\cite{Singh:2002de, Ringwald:2004np, Villaescusa-Navarro:2011loy, LoVerde:2013lta, Elbers:2023mdr, Ichiki:2011ue, Villaescusa-Navarro:2012ilf, LoVerde:2016ahu, deSalas:2017wtt, Zhang:2017ljh, Mertsch:2019qjv, Holm:2023rml, Zimmer:2023jbb, Holm:2023rml, Holm:2024zpr}. The primary goal of most of these studies has been to determine the local neutrino overdensity to make predictions for neutrino direct-detection experiments, such as PTOLEMY~\cite{PTOLEMY:2019hkd}. The motivation for our particular work is a recent paper~\cite{Hotinli:2023scz} that showed that neutrino overdensities around dark-matter halos might produce a detectable contribution to the gravitational lensing of the CMB.  The estimates made in Ref.~\cite{Hotinli:2023scz} relied on a linear approximation~\cite{Singh:2002de} to obtain the radial density profiles of neutrinos around dark-matter halos, due to the computational challenges of numerically evaluating the neutrino-density profiles for large numbers of halos. Such a linear approach, however, is known to underestimate both the true neutrino density~\cite{Ringwald:2004np, Hannestad:2009xu, Tully:2021key} and the corresponding lensing signal. This motivates the development of a fast and efficient code capable of accurately computing non-linear neutrino density profiles.

In this paper, we present a calculation of the neutrino overdensities that arise around spherically symmetric dark-matter halos. The calculation employs an $N$-1--body approach, as in Refs.~\cite{Ringwald:2004np,Brandbyge:2010ge}, to evolve the neutrino phase-space distribution, but integrates trajectories backwards in time~\cite{Mertsch:2019qjv} rather than forwards. This allows for more efficient computation, as only relevant trajectories are integrated. We take dark-matter halos to follow an NFW profile~\cite{Navarro:1996gj} at some observed redshift, having evolved (in a simply and clearly prescribed way) from a homogeneous early Universe. We provide a code written in \texttt{JAX}, called {\tt Cheetah}\footnote{\href{https://github.com/NNSSA/Cheetah}{https://github.com/NNSSA/Cheetah}}, that enables rapid evaluations of neutrino-density profiles around such dark-matter halos, and supports parameterization by neutrino mass, dark-matter halo mass, halo-concentration parameter (which can be put in by hand or inferred from the redshift), and observed redshift. The code generates radial profiles swiftly, typically within seconds on GPUs to minutes on CPUs. Additionally, we provide a \texttt{C} version for use in environments where \texttt{JAX} may not be available or when rapid evaluations for single halos are preferred, alongside a {\tt Python} wrapper. Our framework and accompanying code offer the flexibility to examine the dependency of the results on various assumptions about the halo model, such as the dark matter radial profile and its evolution, as well as extensions to the standard cosmological scenario. We investigate some of these extensions, such as an evolving gravitational constant and nonstandard neutrino phase-space distributions. Finally, besides neutrinos, our method can also be applied to light dark matter components, such as bosonic or fermionic freeze-in dark matter~\cite{Hall:2009bx, Dvorkin:2019zdi, Dvorkin:2022jyg, Heeba:2023bik}. Their massive, non-relativistic behavior at late times offers an interesting extension to our framework for neutrino clustering, which we will explore further in this work.

The paper is organized as follows: In Sec.~\ref{sec:calculations}, we outline our methodology for modeling neutrino trajectories around dark-matter halos, along with the implementation in \texttt{Cheetah}. Numerical results within the standard cosmological scenario are presented in Sec.~\ref{sec:results}. In Sec.~\ref{sec:Num_tests}, we explore the impact of various assumptions on our results, including the evolution of dark-matter halos and new-physics scenarios. Finally, in Sec.~\ref{sec:conclusion}, we provide concluding remarks and discuss some possible future steps. The appendices contain further technical details on the halo parameterization and our numerical implementation.

Throughout this work, we fix the cosmological parameters to the Planck 2018 best-fits: $\Omega_\mathrm{m} = 0.315$ and $h = 0.68$~\cite{Planck:2018vyg}. 

\section{Trajectories and Profiles}
\label{sec:calculations}

\subsection{Equations of Motion}

We begin with a set of definitions that will be used throughout the text. We take the homogeneous background to be a flat FRW Universe with line element
\begin{equation}
    \mathrm{d}s^2 = -\mathrm{d}t^2 + a^2(t) \mathrm{d}\bfx^2 = a^2(\tau) \left[ -\mathrm{d}\tau^2 + \mathrm{d}\bfx^2 \right]\, ,
\end{equation}
where $a(t) = [1+z(t)]^{-1}$ is the scale factor (in terms of redshift $z$), $\tau=\int^t_0 \mathrm{d}t'/a(t')$ is the conformal time, and $\bfx$ is a comoving position.  We take $t_0$, $\tau_0$, and $a_0=1$ to be the values today. The comoving position can be translated into a physical one using $\bfx_\mathrm{p}= a \bfx$. Finally, the expansion rate is given by $H(z)= H_0 \left[\Omega_\mathrm{m} (1+z)^3 + (1-\Omega_\mathrm{m}) \right]^{1/2}$.

In the presence of some density perturbation $\delta\rho(\bfx,\tau)=\rho(\bfx,\tau)-\bar\rho(\tau)$, with $\bar\rho(\tau)$ the mean density, the trajectory of a particle is described as follows (in comoving coordinates)~\cite{Bertschinger:1993xt}:
\begin{equation}
     \ddot \bfx + \frac{\dot a}{a} \dot \bfx + \bfnabla \Phi(\bfx,\tau)=0\, ,
\end{equation}
where the dot denotes a derivative with respect to $\tau$, and the potential $\Phi(\bf x,\tau)$ satisfies the Poisson equation:
\begin{equation}
    \label{eq:Poisson_equation}
    \nabla^2 \Phi(\bfx,\tau) = -4\pi G a^2(\tau) \delta\rho(\bfx,\tau)\, .
\end{equation}
If we restrict our attention to spherically symmetric perturbations, then integration of Eq.~\eqref{eq:Poisson_equation} gives:
\begin{equation}
    \label{eq:potential_derivative}
   \frac{\partial\Phi(r,z)}{\partial r} = -\frac{4 \pi G a^2}{r^2} \int_0^r \mathrm{d}r'\, r'^2  \, \delta\rho(r',z)\ ,
\end{equation}
where $r=|\bfx|$ is the {\it comoving} radius.  Note that we are choosing the origin of coordinates to be the center of the perturbation.

The trajectory of each particle can be determined as follows: We define the peculiar velocity to be $\bfu \equiv \mathrm{d}\bfx/\mathrm{d}\tau$ ($=\mathrm{d}\bfx_\mathrm{p}/\mathrm{d}t$) and comoving velocity $\bfv = a \bfu=\bfu/(1+z)$, and take the motion to be in the $x$-$y$ plane. Using $\mathrm{d}\tau/\mathrm{d}z = [H(z)]^{-1}$, the equations of motion can then be written in terms of four first-order differential equations:
\begin{equation}
    \label{eqn:eoms}
     \frac{\mathrm{d}\bfx}{\mathrm{d}z} = -\frac{\bfv (1+z)}{H(z)}\, , \qquad \frac{\mathrm{d}\bfv}{\mathrm{d}z} = \frac{\bfnabla\Phi}{(1+z) H(z)}\ ,
\end{equation}
for $\bfx = (x,y)$ and similar for $\bfv$, with $\bfnabla\Phi = (\partial\Phi/\partial r) (\bfx/r)$.  Note, as a consistency check, that the peculiar velocity scales as $\bfu \propto (1+z)$ and comoving velocity is constant if $\bfnabla\Phi=0$, as it should.  

For completeness (and comparison with prior work), we note that, with spherical symmetry, particle trajectories can also be written as solutions to the equations
\begin{equation}
     \ddot r + \frac{\dot a}{a} r  -\frac{\ell^2}{2  m_\nu r^3} + \frac{\partial \Phi}{\partial r}=0\, , \qquad \dot\theta = \frac{\ell}{a m_\nu r^2}\, ,
\end{equation}
where $r=(x^2+y^2)^{1/2}$ is the comoving radius, $\theta={\rm arctan}(y/x)$, $\ell=\mathrm{constant}$ is the angular momentum, and $m_\nu$ is the neutrino mass. The system of four differential equations in Eq.~\eqref{eqn:eoms} is in this way, using angular-momentum conservation, reduced to two differential equations. However, we found that this implementation actually increased the computation time, a result of sharp turns near $r=0$ with low-angular-momentum orbits.

\begin{figure*}[t!]
    \centering
    \resizebox{0.85\textwidth}{!}{\includegraphics{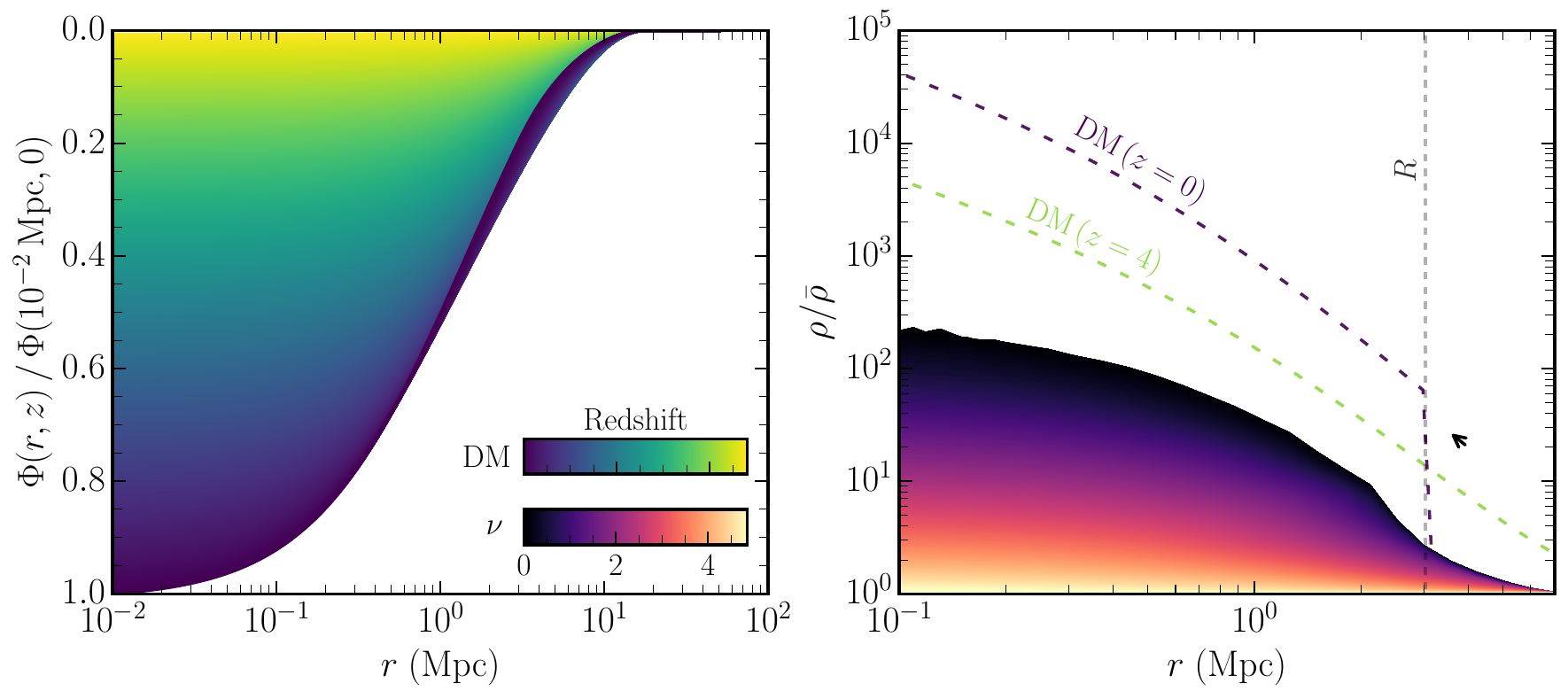}}
    \caption{\textbf{Left:} Evolution of the dark-matter gravitational potential within our halo model. \textbf{Right:} Evolution of the resulting neutrino radial density profiles for $m_{\nu} = 0.3\,\mathrm{eV}$ and $M_\mathrm{h} = 10^{15} \, M_\odot$. The dashed curves represent the dark-matter-halo profile at $z=4$ and $z=0$, with $R$ depicting the comoving virial radius, highlighting the more diffuse nature of neutrino halos.}
    \label{fig:all_potentials}
\end{figure*}

\subsection{Halo Model}

Before solving the equations of motion in Eq.~\eqref{eqn:eoms}, we first need to model the profile and evolution of dark-matter halos, and from that, derive an expression for the derivative of the gravitational potential using Eq.~\eqref{eq:potential_derivative}. We parameterize halos in terms of a mass $M_\mathrm{h}$, redshift $z_0$ at which they are observed (set to $z_0 = 0$), and concentration parameter $c$. The time evolution of the dark-matter mass distribution is modeled as follows: We assume that at $z_0$ a spherical dark-matter halo is virialized to 200 times the mean cosmic density. From this, we can derive the initial redshift $z_i$ at which collapse started using $(1+z_i)/(1+z_0)=200^{1/3}\simeq 5.85$. The halo spans a physical radius $r_{200}=R/(1+z_i)$, with $R$ the comoving radius, and its mass is given by $M_\mathrm{h}=(4\pi/3) r_{200}^3\bar\rho(z_i) = (4\pi/3) R^3 \bar\rho(z_0)$. This region that forms the halo breaks away from the expansion and \underline{re-arranges} itself into the following density distribution (as a function of {\it physical} radius):
\begin{equation}
    \rho_{\mathrm{tot}}(r_\mathrm{p},z) = \begin{cases}
        \begin{aligned}
            &\rho_{\mathrm{NFW}}(r_\mathrm{p},z_0) \xi(z) \\
            &+ \bar{\rho}(z) [1 - \xi(z)]
        \end{aligned}
        & \text{for } r_\mathrm{p} < r_{200} \\[2em] 
        \bar{\rho}(z)[1 - \xi(z)] 
        & \text{for } r_{200} < r_\mathrm{p} < \frac{R}{1+z}
    \end{cases}\, ,
    \label{eqn:rhototal}
\end{equation}
where $\rho_{\rm NFW}(r_\mathrm{p},z_0)$ is the halo density profile at $z_0$, here assumed to be the NFW profile~\cite{Navarro:1996gj}:
\begin{equation}
    \rho_{\rm NFW}(r_\mathrm{p}, z_0) = \frac{\rho_\mathrm{s}}{(r_\mathrm{p}/r_\mathrm{s})(1+ r_\mathrm{p}/r_\mathrm{s})^2}\, ,
\label{eqn:nfw}    
\end{equation}
with $r_\mathrm{s}=r_{200}/c$, $\rho_\mathrm{s} = M_\mathrm{h}  \left[4\pi r_\mathrm{s}^3 I(c) \right]^{-1}$, and $I(c)\equiv \ln(1+c) -c/(1+c)$. Here, $\xi(z)$ is a modulation function that has limits $\xi(z)\to 0$ at $z\to z_i$ and $\xi(z)\to 1$ at $z\to z_0$. A simple choice for $\xi(z)$ is $\xi(z)=(z_i-z)/(z_i-z_0)$ (see Sec.~\ref{sec:Num_tests} for a discussion of this assumption). Eq.~\eqref{eqn:rhototal} describes a halo with a fixed physical radius and mass that is smoothly evolving from a uniform mass distribution at the initial time $z_i$ to an NFW halo profile at $z_0$. Note that the region at radii $r_{200}< r_\mathrm{p}<R/(1+z)$ is underdense in order to compensate for the overdense, NFW profile enclosed within $r_{200}$. Therefore, the total mass enclosed within a comoving radius $R/(1+z)$ is always conserved and is equal to the mean mass enclosed within this radius. That is, the total mass perturbation is thus zero, consistent with how halos are presumed to form through gravitational collapse, rather than simply being added to an otherwise homogeneous Universe. 

With the density profile in Eq.~\eqref{eqn:rhototal}, we are now able to compute the derivative of the gravitational potential as in Eq.~\eqref{eq:potential_derivative}:
\begin{align}
    \frac{\partial\Phi(r,z)}{\partial r} = -\frac{4\pi G }{a r^2} \int_0^{r/(1+z)}\mathrm{d}r_\mathrm{p}\,r_\mathrm{p}^2 \left[ \rho_{\rm tot}(r_\mathrm{p},z) - \bar \rho(z) \right]\, .
\end{align}
An analytical expression for this integral is provided in App.~\ref{app:potentials}. Note that, within the $N$-1--body framework, the contribution of neutrinos to this potential is neglected. We show the evolution of the gravitational potential in the left panel of Fig.~\ref{fig:all_potentials}. With this expression at hand, we can now plug it into the equations of motions in Eq.~\eqref{eqn:eoms} and solve for the neutrino trajectories.

\begin{figure*}[t]
    \centering
    \resizebox{0.83\textwidth}{!}{\includegraphics{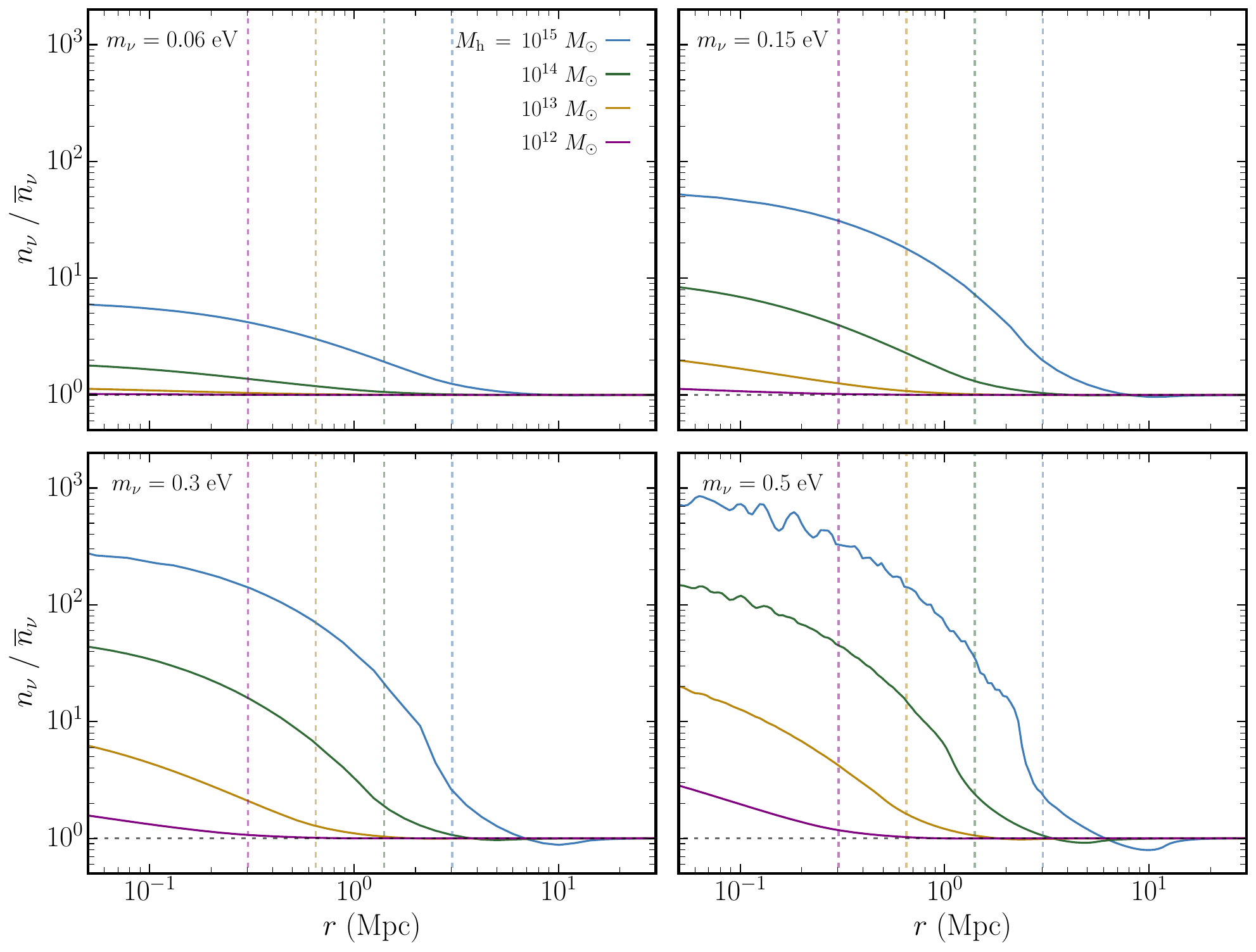}}
    \caption{Dependency of neutrino radial density profiles at $z=0$ on neutrino and dark-matter-halo mass. The vertical dashed lines mark the comoving virial radius of each dark-matter halo, beyond which the neutrino profiles return to the background value. For the heaviest neutrino and dark-matter-halo masses, the neutrino profiles show a slight underdensity at large radii, resulting from mass conservation in our halo model.}
    \label{fig:radial_overdensity}
\end{figure*}

\subsection{Neutrino Profiles}

The equations of motion in Eq.~\eqref{eqn:eoms} can be solved forward in time, as done in, e.g., Ref.~\cite{Ringwald:2004np}. This approach, however, has the drawback that the final position of a neutrino at $z_0$ is often far from the dark-matter halo, requiring many iterative runs to sample enough particles that end up near the halo. A backward method, first proposed in Ref.~\cite{Mertsch:2019qjv}, bypasses such iterative runs entirely by tracing particles back in time, focusing only on those that reach a specified radius. 

We will use this backward method and consider neutrinos that have a comoving position $\bfx_0=(x_0,0,0)$ and comoving velocity $\bfv_0=(v_0\cos\psi,v_0\sin\psi,0)$ at $z_0$. The quantities $x_0$, $v_0$, and $\psi$ are sampled over grids as discussed in Sec.~\ref{subsec:numerical_methods}. These are the initial conditions for the equations of motions in Eq.~\eqref{eqn:eoms}, which we then integrate up to redshift $z_i$. The result of solving these equations is a mapping between comoving momenta at $z_0$ and $z_i$.

This mapping can then be used when computing the neutrino density at $\bfx_0$, which is obtained by integrating the distribution $f_0(x_0,\bfP_0)$ over comoving momenta $\bfP_0$,
\begin{align}
    \label{eq:density_integral}
    n_{\nu,0}(\bfx_0) &= g_\nu \int \frac{\mathrm{d}^3P_0}{(2\pi)^3} f_0(x_0,\mathbf{P}_0) \nonumber \\
    &= \frac{g_\nu}{4\pi^2} \int_{-1}^1 \, \mathrm{d}\mu \, \int_0^\infty \mathrm{d}P_0 P_0^2 f_0(x_0,P_0,\mu)\, ,
\end{align}
where $\mu = \cos\psi$. The neutrino distribution function in the homogeneous Universe, before halo formation, is the Fermi-Dirac distribution $f_{\rm FD}(P) = \left[1 + {\rm exp}(P/T_{\nu,0} \right]$, with $T_{\nu,0}=1.95\,\mathrm{K}$ the neutrino temperature today. We use Liouville's theorem~\cite{2008gady.book.....B}, which tells us that the phase-space density along the computed neutrino trajectories is conserved, to identify
\begin{equation}
     f_0(x_0,P_0,\mu) = f_{\rm FD}\left[P_i(x_0,P_0,\mu) \right]\, ,
\end{equation}
where $P_i$ is the magnitude of the comoving momentum at $z_i$. Eq.~\eqref{eq:density_integral} can then be rewritten as
\begin{equation}
    n_{\nu,0}(\bfx_0) = \frac{g_\nu}{4\pi^2} \int_{-1}^1\, \mathrm{d}\mu\, \int_0^\infty \mathrm{d}P_0 P_0^2\, f_{\rm FD}\left[ P_i(x_0,P_0,\mu)\right]\, ,
\label{eqn:localdensity}    
\end{equation}
which should be compared to the background density,
\begin{equation}
    \bar n_{\nu} = \frac{g_\nu}{2\pi^2}   \int_0^\infty \mathrm{d}PP^2f_{\rm FD}(P)\, , 
\end{equation}
to obtain the neutrino overdensity. For convenience, we will omit the subscript 0 in $n_{\nu,0}$ moving forward.

\begin{figure*}[t]
    \centering
    \includegraphics[width=\textwidth]{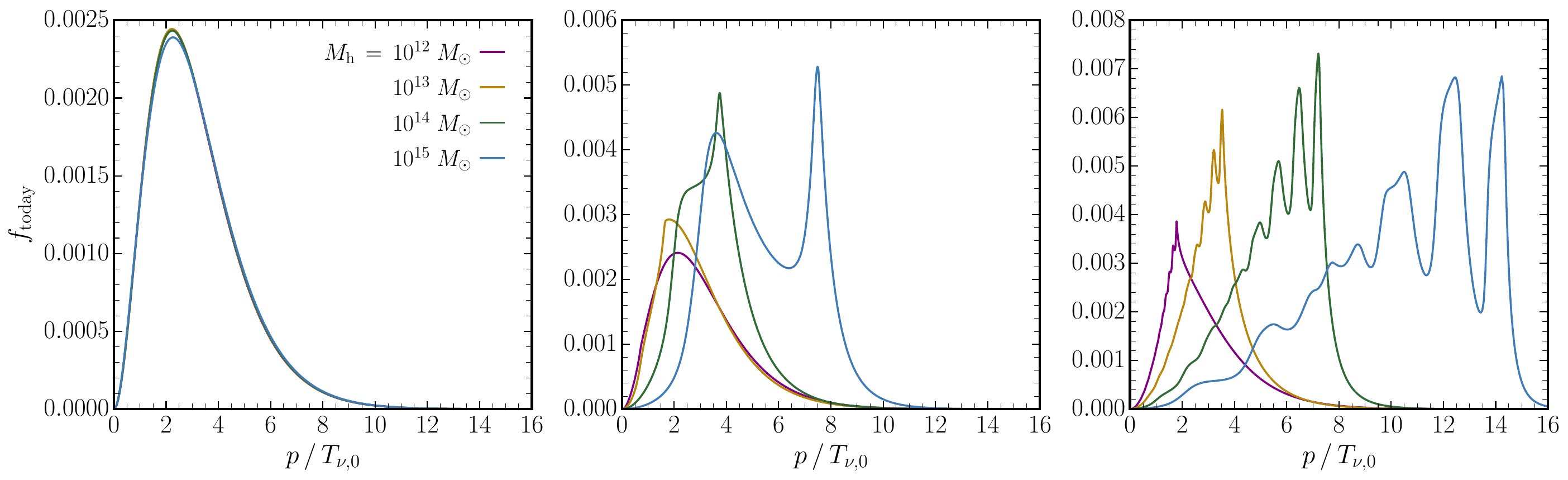}
    \caption{Neutrino phase-space distributions at large radii $r\gg r_{200}$ (\textbf{left}), the virial radius $r_{200}$ (\textbf{middle}), and the scale radius $r_\mathrm{s}$ (\textbf{right}). Here, the redshift and neutrino mass are set equal to 0 and $0.3\,\mathrm{eV}$, respectively. At large radii, the distributions follow the Fermi-Dirac distribution, while at smaller radii they exhibit clustered peaks, due to the existence of distinct phase-space shells. These peaks sharpen for colder neutrinos (large $m_\nu$) and broaden for warmer neutrinos (small $m_\nu$).}
    \label{fig:integrands}
\end{figure*}

\subsection{Numerical Methods}
\label{subsec:numerical_methods}
For each position $\bfx_0$, we need to evaluate the two-dimensional integral in Eq.~\eqref{eqn:localdensity} over the neutrino phase-space distribution, which itself is derived from solving the four coupled differential equations in Eq.~\eqref{eqn:eoms}. The differential equations for the trajectories are, however, fairly simple and rapidly evaluated in parallel in \texttt{Cheetah}. We aim for results with ${\sim}1\%$ precision, which can be achieved without dense sampling of the integrand.

The equations of motion in Eq.~\eqref{eqn:eoms} are independent of mass, allowing the same trajectories to be used for different neutrino masses. {\tt Cheetah} solves these equations for a chosen fiducial mass (here, $m_\nu = 0.3\,\mathrm{eV}$), and the resulting velocities are scaled by a grid of neutrino masses to obtain the corresponding momenta. Properly selecting the initial grid of momenta is essential for accurately sampling the phase-space distribution, as confirmed by inspecting the code's integrand outputs.

We initialize neutrino trajectories using parameter grids with 10 linearly-spaced values for $\mu$ in the range $[0,1]$, 20 logarithmically-spaced values for $x_0$ in the range $[10\,\mathrm{kpc}, 50\,\mathrm{Mpc}]$, and 15 linearly-spaced values for the neutrino mass in the range $[0.01, 0.5]\,\mathrm{eV}$. In addition, we have 200 momenta values in the range [0.01,400]$T_{\nu,0}$, and velocities are determined through dividing by the neutrino mass. This means that for each $x_0$, we solve the equations in Eq.~\eqref{eqn:eoms} 2000 times. The architecture of \texttt{Cheetah} is optimized for GPU-based parallelization of $N$-1--body runs, making it ideal for HPC environments.

In addition to the \texttt{JAX} code, we provide a \texttt{C} implementation that calculates neutrino overdensities for one neutrino mass at a time. This version is ideal for users without access to HPC resources or when \texttt{JAX} is unavailable. The \texttt{C} code follows the same approach as discussed above, but includes several approximations, some of which we detail in App.~\ref{app:numericalintegrations}, allowing it to achieve run times of just a few seconds on a laptop CPU.

\begin{figure*}[t!]
    \centering
    \includegraphics[width=\textwidth]{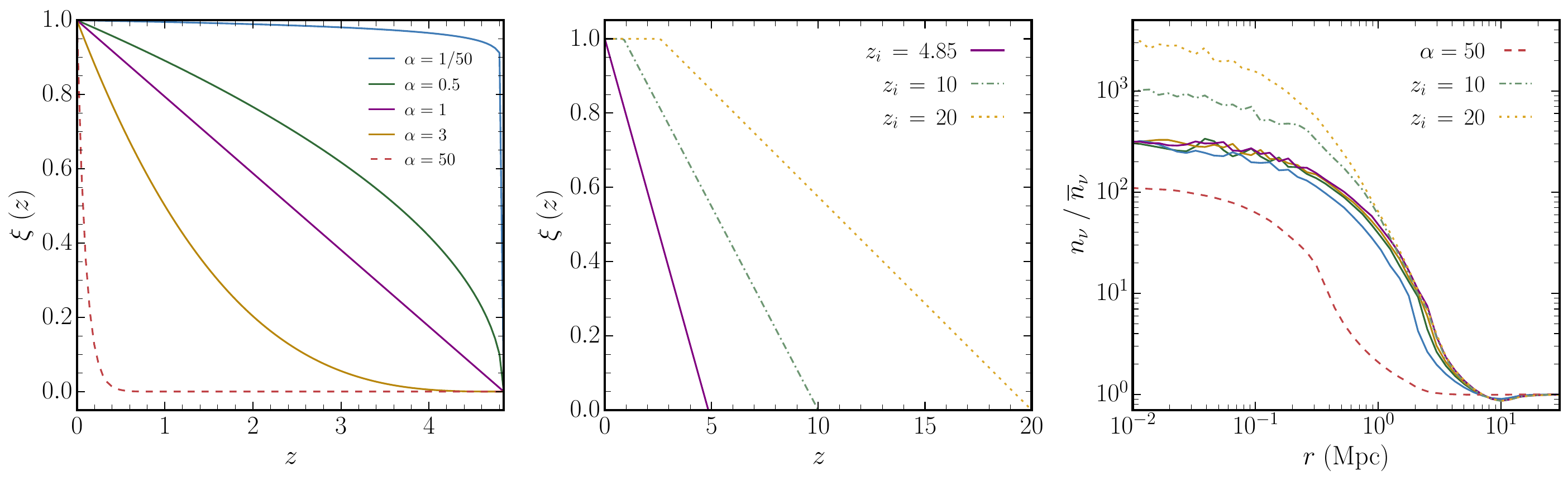}
    \caption{Impact of different dark-matter-halo formation histories on neutrino density profiles. \textbf{Left:} Deviations from linear growth in the modulation function governing the dark-matter-halo growth rate. \textbf{Middle:} Modulation functions with a different initial redshift. \textbf{Right:} The resulting neutrino density profiles at $z=0$. The effect of these modulations on the final neutrino profile is minimal in most cases, except for a suppression with rapid late-time growth and an enhancement when dark-matter halo formation begins early.}
    \label{fig:combined_mod_func}
\end{figure*}

\section{Results}\label{sec:results}
Having outlined our methodology, we can now run our pipeline to obtain neutrino overdensity profiles along with their redshift evolution. We present these in the right panel of Fig.~\ref{fig:all_potentials}. As noted in Ref.~\cite{Hotinli:2023scz}, the neutrino profiles are significantly more diffuse than those of the dark matter, exhibiting a smooth inner core and extending beyond the dark-matter halo (as indicated by the vertical R line). Additionally, we find that most of the neutrino halo growth occurs at late times, making it less sensitive to the details of dark-matter-halo formation at earlier times (which we further elaborate on in Sec.~\ref{sec:Num_tests}).

In Fig.~\ref{fig:radial_overdensity}, we show the dependence of neutrino profiles on both neutrino mass and dark-matter-halo mass. As expected, heavier neutrinos or dark-matter halos lead to more significant clustering. As can be seen in the bottom plots, the profiles exhibit an underdensity between $r_{200}$ and R/(1+$z_i$), with the underdensity becoming more pronounced when the central clustering is greater. Shortly beyond R/(1+$z_i$), the densities return to the background value.

We also examine the neutrino phase-space distribution at different radii, see Fig.~\ref{fig:integrands}. At large radii (left panel), the distribution follows the original Fermi-Dirac profile. However, at smaller radii (middle and right panels), multiple peaks emerge, especially for higher dark-matter-halo masses. These peaks result from the different phase-space shells in the spherical-infall model -- at a given radius, there are neutrinos that have just fallen in, as well as those that have oscillated about the origin one, two, or more times. In the limit of cold neutrinos (larger $m_\nu$), these peaks become sharper (and identical to those in the dark-matter velocity distribution), while for warmer neutrinos (smaller $m_\nu$), they broaden and merge.

At a given radial position, we find that there exists an approximate power-law relationship between neutrino mass and neutrino overdensity, with a power-law index of $2.5$ for $M_\mathrm{h}$ = $10^{15} \, \text{M}_{\odot}$ and $2$ for $M_\mathrm{h}$ = $10^{12} \, \text{M}_{\odot}$\footnote{We provide a plot on the \href{https://github.com/NNSSA/Cheetah/tree/main}{GitHub} page of this project.}. Although our dark-matter-halo model differs from those in previous works, our computed neutrino profiles are largely consistent with those reported in~\cite{Ringwald:2004np, Zhang:2017ljh, Mertsch:2019qjv}, across both radial positions and neutrino masses.

\section{Numerical Tests}\label{sec:Num_tests}

\subsection{Dark-matter Halo Formation}
\label{subsec:dm_halo_formation}
In our halo model, the growth rate of dark-matter halos is governed by the modulation function $\xi(z)$ in Eq.~(\ref{eqn:rhototal}). A simple parameterization is given by $\xi(z) = [(z_i-z)/(z_i-z_0)]^\alpha$, where we have so far used a default value of $\alpha=1$ for the power-law index, corresponding to linear growth with redshift. In contrast, higher values of $\alpha$ emphasize late-time growth, while lower values emphasize early-time growth. We explore the impact of different $\alpha$ values and show the results in Fig.~\ref{fig:combined_mod_func}. The left panel depicts the modulation function for different $\alpha$, including some extreme cases, and the right panel shows the effect on the neutrino profiles. Our findings indicate that the dark-matter-halo growth has minimal influence on the final neutrino density profile, except for a suppression in an extreme case where the halo grows very rapidly at late times.

To further test the dependence of our results on the assumed halo evolution, we consider a scenario where halo growth begins and ends earlier than in the base case, see middle panel in Fig.~\ref{fig:combined_mod_func}. As shown, the earlier onset of dark-matter-halo growth leads to higher neutrino densities at later times (see right panel), as neutrinos have more time to cluster around the virialized halo. 

Another important factor is the halo-concentration parameter, which evolves with redshift and depends on halo mass~\cite{Ludlow:2012wh, Ludlow:2013vxa, Dutton:2014xda, Diemer:2014gba, Groener:2015cxa, Correa:2015dva}. Due to the analytical nature of our halo model, there is some uncertainty regarding the appropriate value for this parameter. In \texttt{Cheetah}, we allow for two possibilities: One where the concentration parameter evolves with redshift for a given dark-matter-halo mass using the parameterization from Ref.~\cite{Correa:2015dva}, and another where it remains constant. We find that, as long as the concentration parameter at $z_0$ is kept roughly the same, the neutrino density profile is largely unaffected. This is likely due to our earlier observation that the late-time stages of halo growth are most relevant, making the exact evolution of the concentration parameter less significant for neutrino density profiles.

\begin{figure*}[!t]
    \centering
    \includegraphics[width=0.87\textwidth]{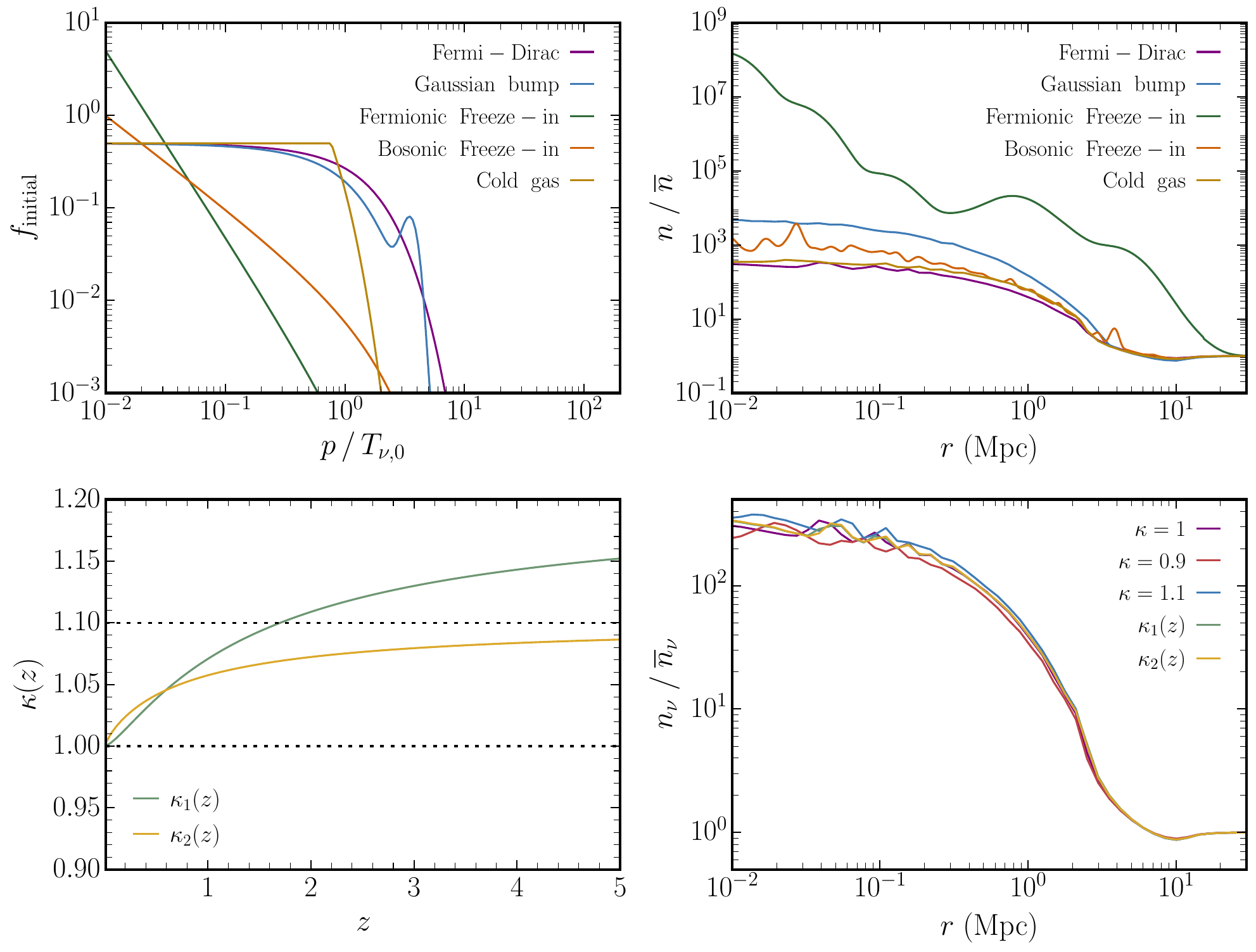}
    \caption{Illustration of the impact of new physics on the clustering of neutrinos and non-cold relics (see surrounding text for details). We fix the particle mass at $m = 0.3\, \mathrm{eV}$ and the dark-matter-halo mass at $10^{15} \, M_\odot$. \textbf{Top panels:} The effect of alternative distribution functions beyond Fermi-Dirac (top left) on the resulting profiles (top right). \textbf{Bottom panels:} Variations in the gravitational constant via the modulation function $\kappa$ (bottom left) and their impact on neutrino profiles (bottom right). We used $\alpha$ and $\beta$ values of 0.2 and 1.5 for $\kappa_1$, and 0.1 and 0.8 for $\kappa_2$.}
    \label{fig:new_physics}
\end{figure*}

\subsection{New Physics}
\label{subsec:new_physics}
In the standard $\Lambda$CDM model, relic neutrinos follow a Fermi-Dirac distribution with a present-day temperature of $1.95\,\mathrm{K}$. While current CMB anisotropy measurements can be used to constrain neutrino energy densities, they are insensitive to specific features in the neutrino distribution~\cite{Cuoco:2005qr, Oldengott:2019lke, Alvey:2021xmq}. Here, we examine the effects of spectral distortions on the clustering of neutrinos by exploring different initial distributions, see top-left panel in Fig.~\ref{fig:new_physics}. Specifically, we consider a thermal distribution with a lower temperature and a Gaussian bump at higher momenta~\cite{Alvey:2021xmq}, and a cold Fermi gas distribution inspired by Ref.~\cite{Ferrer:1999ad}, which decays more rapidly than the Fermi-Dirac at higher momenta. The resulting neutrino profiles are shown in the top-right panel, where, as expected, enhancements at low momenta lead to an increased clustering, while suppression reduces it. 

In addition to neutrinos, we examine the clustering of other non-cold relics with different distribution functions. In particular, we explore a freeze-in dark matter component~\cite{Dvorkin:2019zdi,DePorzio:2020wcz, Xu:2021rwg} that could have either a fermionic or bosonic phase-space distribution, see top-left panel of Fig.~\ref{fig:new_physics}. Similar to the low-momentum deviations from the Fermi-Dirac distribution for neutrinos, these low-momentum enhancements contribute to an increased clustering, emphasizing their potential for distinct observational signatures.

Besides changes in the distribution function, we also explore small deviations from General Relativity by altering Newton's constant with a multiplicative factor, $\kappa$, such that the gravitational potential now includes $\kappa\times G$. We consider two scenarios: In one, $\kappa$ has a time dependence and decays to unity at late times, $\kappa$ = $1 + \alpha  \left[z/(1+z)\right]^{\beta}$, staying within the constraints from Ref.~\cite{Andrade:2023pws}, while in the other $\kappa$ is fixed (see bottom-left panel of Fig.~\ref{fig:new_physics}). As shown in the bottom-right panel, the neutrino profile is largely insensitive to these variations.

\section{Conclusions and Outlook}
\label{sec:conclusion}

In the previous sections, we presented a framework for modeling and simulating relic-neutrino clustering around dark-matter halos. We employed an analytical approach to describe the evolution of dark-matter halos, whose gravitational potential captures neutrinos, while ensuring the total mass of the system is conserved. The trajectories of relic neutrinos in this potential are described by Hamilton's equations of motion, which we solved using \texttt{Cheetah}, a highly parallelizable code capable of delivering solutions within seconds. 

At its core, \texttt{Cheetah} utilizes the $N$-1--body approach and integrates the equations of motion backward in time to efficiently map current phase-space properties of neutrinos to their past ones. Then, by applying Liouville's theorem, the present-day phase-space distribution of neutrinos can be derived from the high-redshift distribution, which in our case is the Fermi-Dirac distribution. Finally, after integrating this distribution, we obtained neutrino density profiles, parameterized by neutrino mass, dark-matter halo mass, halo-concentration parameter, and observed redshift.

Our main result is a rapid, self-consistent calculation of neutrino density profiles around dark-matter halos. The framework is also highly adaptable, allowing for easy testing of other non-cold relics or modifications to General Relativity. This makes the code especially useful for future CMB and large-scale structure analyses focused on detecting signals from such relics.

In what follows, we elaborate on the key model assumptions we have made and discuss potential extensions of this work.\\

\noindent \textbf{Halo model.} In our analysis, we assumed dark-matter halos to be spherically symmetric and isolated. While the code can generate neutrino density profiles in the presence of a population of dark-matter halos, it does not explicitly account for interactions between halos or their cumulative effects. As noted in Ref.~\cite{Zimmer:2023jbb}, an asymmetric dark-matter velocity distribution can lead to a depletion of lower-momentum neutrino states due to intervening material, leaving clustered neutrinos with momenta below the escape velocity but above this depletion threshold. However, the tests shown in Fig.~\ref{fig:combined_mod_func} indicate that the final neutrino profile is largely independent of the precise dark-matter formation history, but rather depends mostly on when halo formation started. Although {\tt Cheetah} cannot effectively probe the depletion of lower momentum states caused by intervening halos and material, this result suggests minimal change in the final neutrino clustering. The inclusion of multiple halos in the system would nonetheless be an interesting extension of this work, and particularly relevant for producing maps of relic-neutrino anisotropies~\cite{Zimmer:2023jbb, Zimmer:2024max}.\\

\noindent \textbf{Baryons.} Baryons too contribute to the gravitational potential well and will increase the clustering of neutrinos and alter the subsequent observational signature. In contrast to dark matter, baryons follow different profiles within halos that are rather non-universal~\cite{Oman:2015xda, McCarthy:2016mry}. As such, a useful approach here can be to model these using empirical methods or semi-analytical prescriptions that allow to marginalize over unknown parameters~\cite{Berlind:2001xk,Harikane:2015unm, Leauthaud:2011gw, Sorini:2021eac}.\\

\noindent \textbf{Neutrino contribution to potential.} In our calculation, the contribution of neutrinos to the gravitational potential was neglected, though a small enhancement in clustering is expected if this were included. Since most of the neutrino-halo growth occurs at late redshifts and the neutrino masses are relatively small, their contribution to the gravitational potential is minimal, especially when compared to that of the dark matter. To account for this effect numerically, one can iteratively compute the neutrino density profile, adding an additional potential term at each time step until convergence is achieved.\\

\noindent \textbf{CMB lensing.} As demonstrated in Ref.~\cite{Hotinli:2023scz}, neutrino halos induce a lensing signal in the CMB that may become detectable with next-generation experiments. This signal probes the overall matter distribution, and has thus contributions from dark matter, baryons, and neutrinos that may be degenerate with one another. The baryonic component can be isolated using, for instance, the thermal Sunyaev-Zeldovich effect~\cite{ACT:2023wcq, SimonsObservatory:2018koc}. The neutrino signature can then be probed by marginalizing over the parameters of both neutrino and dark-matter halos. \\

\noindent \textbf{Other dark-matter models.} The flexibility of \texttt{Cheetah} allows for the incorporation of various dark matter models to study their effects on neutrino clustering. For example, a warm dark matter model can be implemented by introducing a parametric function that suppresses small-scale power in the matter power spectrum~\cite{Lesgourgues:2011rh, Ludlow:2016ifl, Lovell:2013ola}. Alternative models, such as decaying dark matter, introduce a time-dependent modification to gravitational potential wells, affecting neutrino clustering at different stages of cosmic history~\cite{Chluba:2011hw, Bernal:2017kxu, Wang:2014ina, Chen:2003gz, Polisensky:2010rw}. These different properties of dark matter can have notable consequences for the resulting neutrino profiles. \\

\noindent \textbf{Non-linear regime.} The impact of massive neutrinos in linear theory has been extensively studied, with the primary effect being the suppression of the matter power spectrum at small scales due to their free-streaming behavior~\cite{Lesgourgues:2006nd, Agarwal:2010mt, Bashinsky:2003tk}. However, at the nonlinear level, incorporating the effects of neutrinos on large-scale structure is significantly more challenging, as their initial phase-space distribution is six-dimensional, unlike the three-dimensional distribution of cold dark matter. The framework established here can provide a valuable testbed for evaluating various techniques that have been developed to approximate the effects of massive neutrinos on nonlinear structure formation~\cite{Brandbyge:2008rv, Bird:2011rb, AliHaimoud:2012fzp, Inman:2016qmg, Banerjee:2016zaa, Bird:2018all, Elbers:2020lbn, Bayer:2021kwg, Banerjee:2022era}.\\

In summary, we have developed an effective framework for simulating neutrino clustering around dark-matter halos. Our aim was to create a versatile and practical tool for studying future observational signatures of relic neutrinos, such as their gravitational lensing effect on the CMB. The rapid evaluation capabilities of our code allow for the efficient exploration of a wide range of theoretical models with minimal computational overhead, making it especially useful for aligning theoretical predictions with emerging observational data.

\begin{acknowledgments}
We thank Bryan Scott and Benjamin Joachimi for useful discussions. This work was supported by NSF Grant No.\ 2412361, the Simons Foundation, and the Templeton Foundation. K.W. thanks the MD Space Grant Consortium and William H. Miller Graduate Fellowships for funding this work. K.W. also thanks the LSST-DA Data Science Fellowship Program, which is funded by LSST-DA, the Brinson Foundation, and the Moore Foundation; his participation in the program has benefited this work. N.S. was supported by a Horizon Fellowship from Johns Hopkins University. This work was carried out at the Advanced Research Computing at Hopkins (ARCH) core facility (arch.jhu.edu), which is supported by the National Science Foundation (NSF) grant number OAC1920103.
\end{acknowledgments}

\bibliography{refs}

\appendix

\onecolumngrid
\section{Expression for Potential} 
\label{app:potentials}
In this appendix, we provide a simple expression for the gravitational potential and its derivative by integrating the Poisson equation in Eq.~\eqref{eq:Poisson_equation} using the density distribution within our halo model (Eq.~\eqref{eqn:rhototal}):
\begin{equation}
\Phi(r, z) = \frac{GM \xi(z)}{a}
\begin{cases}
    -\left[\frac{1}{r I(c)} \log\left(1 + \frac{ar}{r_\mathrm{s}}\right) - \frac{a}{r_{200} I(c)} \log(1 + c) + \frac{a}{r_{200}}\right] + \frac{(3(aR)^2 - r^2)}{2(aR)^3} & \text{for } r_\mathrm{p} < r_{200} \\
    -\frac{1}{r} + \frac{(3(aR)^2 - r^2)}{2(aR)^3} & \text{for $r_{200}<r_\mathrm{p}<R/(1+z)$} \\
    0 & \text{otherwise}
\end{cases}\ ,
\label{eq:phi_rz}
\end{equation}
with $a=1/(1+z)$. This results in the following expression for the gradient of the potential:
\begin{eqnarray}
     \frac{\partial\Phi(r,z)}{\partial r} &=& -\frac{4 \pi G a^2}{r^2} \int_0^r \mathrm{d}r' (r')^2 \delta\rho(r',z) \nonumber \\
     &=& -\frac{4\pi Ga^2}{r^2} \int_0^r \mathrm{d}r' (r')^2 \left[\rho_{\rm tot}\left(\frac{r'}{(1+z)},z\right)-\bar\rho(z) \right] \nonumber \\
     &=& -\frac{4\pi G a^2}{r^2}(1+z)^3 \int_0^{r/(1+z)} \mathrm{d}r_\mathrm{p} r_\mathrm{p}^2 \left[ \rho_{\rm tot}(r_\mathrm{p},z) - \bar \rho(z) \right] \nonumber \\
     & = & -\frac{G (1+z)}{r^2}\xi(z) M_\mathrm{h}
     \begin{cases} \left[\frac{ I\left[(r/r_\mathrm{s})/(1+z)
     \right]}{I(c)} - \left(\frac{r}{ r_{200}  
     (1+z_i)}\right)^3 \right] & \text{for $r_\mathrm{p}<r_{200}$} \\   
      \left[1 - \left(\frac{r}{ r_{200} (1+z_i)}\right)^3 \right] & \text{for $r_{200}
      <r_\mathrm{p}<R/(1+z)$} \\   
      0 & \text{otherwise}
     \end{cases}\ ,
\end{eqnarray}
We see that, at $z > z_i$, $\Phi(r, z) = 0$ and $\partial \Phi(r, z)/\partial r = 0$.

\section{Approximation Scheme}
\label{app:numericalintegrations}
In this appendix, we introduce an approximation scheme that can be implemented to further reduce computational loads. At large radii $r > r_{200}(1+z_i)$, these points are outside the halo region within which $\Phi\neq0$. However, the neutrino density here may be affected by neutrinos whose trajectories have passed through the $\Phi\neq0$ region. Given that, if $\Phi=0$, neutrinos propagate along straight lines in comoving coordinates, the equations of motion in Eq.~\eqref{eqn:eoms} need to be solved only for neutrino trajectories with $\left(1-[r/r_{200}(1+z_i)]^2\right)^{1/2}<\mu<1$, i.e., only those that pass through the region where $\Phi\neq0$. This calculation also only needs to be performed for comoving velocities large enough for the neutrino to travel from $r_{\rm 200}(1+z_i)$ to $r$. A neutrino with comoving velocity $v$ can propagate a distance [cf., Eq.~(\ref{eqn:eoms})],
\begin{equation}
   vI_{io} \equiv v\int_{z_0}^{z_i} \frac{(1+z)\, \mathrm{d}z}{H(z)}\, ,
\end{equation}
between redshifts $z_0$ and $z_i$.  We thus need to solve the equations of motion only for neutrinos with velocities $v_0>(r-r_{200}(1+z_i)/I_{io}$.  For smaller $\mu$ or smaller $v_0$, the momentum is constant. The propagation distance also indicates that neutrinos will feel the effects of the halo at points that are as far as $v_{\rm max}I_{io}$ away from it. The integral must be evaluated numerically, but is of order $\sim v/H_0$ which can be $\sim100\,\mathrm{Mpc}$ for velocities of about $5000\,\mathrm{km/sec}$. This integration scheme is used in the \texttt{C} version of the code.

\end{document}